\documentclass[aps,prl,preprintnumbers,amsmath,amssymb,twocolumn,superscriptaddress,showpacs,floatfix]{revtex4}
\usepackage{bbm}
\usepackage{mathrsfs}
\usepackage{amsmath,amssymb,bm,comment}
\usepackage{hyperref}
\usepackage{graphicx}
\usepackage{dcolumn}
\usepackage{bm}
\usepackage{stmaryrd}
\usepackage{color}
\usepackage{fancyhdr}

\begin{document}

\title{ Helicity polarization in relativistic heavy ion collisions }

\author{Jian-Hua Gao}
\email{gaojh@sdu.edu.cn}
\affiliation{Shandong Provincial Key Laboratory of Optical Astronomy and Solar-Terrestrial Environment,
Institute of Space Sciences, Shandong University, Weihai, Shandong 264209, China}

\begin{abstract}
We discuss the  helicity polarization which  can be locally induced from both vorticity and helicity charge in non-central heavy ion collisions.
Helicity charge redistribution can be generated in viscous fluid and contributes to azimuthal asymmetry of the polarization along global angular momentum or beam momentum. We also discuss on detecting the initial net helicity charge  from topological charge fluctuation or initial color longitudinal field by  the helicity  correlation of  two hyperons  and  the helicity  alignment of vector mesons in central heavy ion collisions.
\end{abstract}


\maketitle

\thispagestyle{fancy}
\renewcommand{\headrulewidth}{0pt}
\section{Introduction}

Spin polarization effect  has drawn much attention in relativistic heavy ion collisions recently. Spin freedom provides us a unique probe to detect the feature of  quark gluon plasma in quantum level.  Much development has been made along this direction  on either  experimental  aspect
\cite{Abelev:2007zk,Abelev:2008ag,STAR:2017ckg,Adam:2018ivw,Adam:2019srw,Acharya:2019ryw,Acharya:2019vpe,Adam:2020pti} or
theoretical aspect\cite{Liang:2004ph,Liang:2004xn,Betz:2007kg,Gao:2007bc,Huang:2011ru,Becattini:2013fla,Becattini:2013vja,
Xie:2015xpa,Pang:2016igs,Becattini:2016gvu,Karpenko:2016jyx,
Li:2017slc,Sun:2017xhx,Han:2017hdi,Becattini:2017gcx,Yang:2017sdk,Kolomeitsev:2018svb,Xia:2018tes,
Wei:2018zfb,Sun:2018bjl,Florkowski:2019voj,Xia:2019fjf,Wu:2019eyi,Guo:2019mgh,
Liu:2019krs,Sheng:2019kmk,Sheng:2020ghv,Xia:2020tyd,Liang:2019pst,Fu:2020oxj,Wang:2021owk,Liu:2021uhn,
Fu:2021pok,Becattini:2021suc,Becattini:2021iol}. Some relevant review on spin effects in relativistic heavy ion collisions are available in Refs
\cite{Florkowski:2018fap,Becattini:2020ngo,Liu:2020ymh,Gao:2020vbh,Gao:2020lxh,Huang:2020dtn,Becattini:2021lfq}.
Most of these works concentrate on the global or local polarization along the global angular momentum (transverse polarization) or beam momentum (longitudinal polarization) for
hyperons or vector mesons. In this paper,we will discuss another possible spin polarization along the momentum of final hadrons. In order to distinguish such polarization with the longitudinal polarization along the beam momentum, we denote it as helicity polarization.  Other earlier works associated with the helicity  in heavy ion collisions can be found
in Refs.\cite{Jacob:1987sj,Ambrus:2019ayb,Ambrus:2019khr,Ambrus:2020oiw,Becattini:2020xbh}.

\section{Helicity polarization }
In relativistic heavy ion collisions, the single-particle mean spin vector $S^\mu(p)$ is given by\cite{Becattini:2013fla}
\begin{eqnarray}
\label{S-mu}
S^\mu(p) &=&-\frac{1}{8m} \epsilon^{\mu\nu\rho\sigma} p_\nu \frac{\int d\Sigma_\alpha p^\alpha \varpi_{\rho\sigma}  f (1-f)}
{\int d\Sigma_\alpha p^\alpha f },
\end{eqnarray}
where $f$ is Fermion-Dirac distribution function
\begin{eqnarray}
f &=&\frac{1}{e^{\beta_\mu  p^\mu }+1},\ \  \textrm{with}\ \ \ \beta_\mu = \frac{u_\mu}{T}\  \textrm{and}\   u^2=1
\end{eqnarray}
and  $\varpi_{\rho\sigma}$ is thermal vorticity
\begin{eqnarray}
\varpi_{\mu\nu}=-\frac{1}{2}\left(\partial_\mu \beta_\nu -\partial_\nu \beta_\mu\right)\;.
\end{eqnarray}
In this paper, we also need  the temperature  vorticity tensor $\Omega^{\mu\nu}$ defined by
\begin{eqnarray}
\label{TV}
\Omega^{\mu\nu}&=& \partial^\mu \left( T u^\nu\right) -  \partial^\nu \left( T u^\mu\right) ,
\end{eqnarray}
and its dual tensor
\begin{eqnarray}
\label{dualTV}
\tilde \Omega^{\mu\nu}&=&\frac{1}{2} \epsilon^{\mu\nu\rho\sigma}\Omega_{\rho\sigma}.
\end{eqnarray}

In the previous work, the polarization along the global angular momentum $-y$, impact parameter $x$ or the bean momentum $z$  are all investigated. Now let us consider the
polarization along the direction of the particle's momentum --- helicity polarization. We contract the unit  3-vector momentum $\hat {\bf p}$
with Eq.(\ref{S-mu}) and obtain
\begin{eqnarray}
\label{Jmu-1-a}
 S^h &\equiv& \hat{\bf p}\cdot {\bf S}(p) \nonumber\\
&=& \frac{1}{8m}  \frac{\int d\Sigma_\alpha p^\alpha  \hat{\bf p}\cdot \left(\pmb{\nabla}\times \beta \bf u\right)  f (1-f)}
{\int d\Sigma_\alpha p^\alpha  f },
\end{eqnarray}
where the superscript $h$ denotes helicity and   ${\bf u}$ is the spatial component of 4-vector $u^\mu$.
It is very interesting that only the spatial components of vorticity tensor are involved and the time component does not contribute at all.
Hence we can obtain the information of  the pure spatial vorticity by  measuring the helicity polarization if we neglect the helicity charge at the beginning.

Following the assumption used in Ref.\cite{Becattini:2017gcx} that the temperature vorticity vanishes at all times for ideal uncharged fluid and
the temperature on the decoupling hyper-surface only  depends on the Bjorken time, Eq. (\ref{S-mu}) can be reduced into
\begin{eqnarray}
\label{S-mu-2}
S^\mu(p) &=& \frac{1}{4mT} \frac{d T}{d\tau}\frac{p_\nu \partial_\sigma^p \int d\Sigma_\alpha p^\alpha  f  }
{\int d\Sigma_\alpha p^\alpha f }\nonumber\\
& &\times  \left(\epsilon^{\mu\nu 0 \sigma}\cosh\eta - \epsilon^{\mu\nu 3 \sigma}\sinh\eta \right).
\end{eqnarray}
It is obvious that the first term contributes to the longitudinal polarization which has been fully discussed by Becattini and Karpenko in
Ref.\cite{Becattini:2017gcx} and can be approximated at rapidity $Y=0$ as
\begin{eqnarray}
\label{S-mu-2}
 {S}^z &\approx&\frac{1}{mT} \frac{dT}{d\tau} v_2(p_T) \sin 2\phi,
\end{eqnarray}
while the second term contributes to the helicity polarization and can be approximated  as
\begin{eqnarray}
\label{S-mu-2}
 S^h &\approx&\frac{Y}{mT}  \frac{dT}{d\tau} v_2(p_T) \sin 2\phi =Y S^z,
\end{eqnarray}
at small rapidity.  We note that  the helicity polarization must be rapidity odd and  vanish at $Y=0$ which is different from the longitudinal polarization.
Hence we should detect the helicity polarization  locally with $Y>0$ or $Y<0$ separately.

\section{Helicity charge redistribution}
The helicity charge could contribute to helicity polarization as well. For simplicity, we will restrict ourselves to the chiral limit so that we  can use the chiral kinetic theory to
deal with it. At chiral limit, the helicity coincides  with the chirality except a trivial opposite sign for antiparticles. In the chiral kinetic theory, the polarization vector of fermion particle is proportional to the axial Wigner function. In a chiral system with free fermions  in local equilibrium, the axial Wigner function is given by \cite{Gao:2017gfq}
\begin{eqnarray}
\label{Jmu-1-a}
\mathcal{A}^\mu(x,p) = \frac{p^\mu}{E_p} \sum_\lambda \lambda f_\lambda - \frac{\epsilon^{\mu\nu\rho\sigma}p_\nu \varpi_{\rho\sigma}}{4E_p}\sum_\lambda f_\lambda (1-f_\lambda),\ \
\end{eqnarray}
where $E_p=|\bf{p}|$ and $\lambda=\pm 1$ denotes the  particle's  helicity or chirality  with righthand ($\lambda=+1$) or lefthand ($\lambda=-1$). Here $f_\lambda$ represents the Fermi-Dirac distribution
with helicity $\lambda$
\begin{eqnarray}
\label{f-s}
f_\lambda &=&\frac{1}{e^{\beta_\mu  p^\mu -\lambda\bar\mu_5}+1},
\end{eqnarray}
where  $\bar \mu_5 = \mu_5/T$ with $\mu_5$ being axial chemical potential. When there is no axial charge with $\mu_5=0$, the first term will vanish and the second term will give rise to the polarization in Eq.(\ref{S-mu}). However the chiral separate effect or local polarization effect  from the vorticity \cite{Gao:2012ix} can induce a redistribution of  axial charge and initial zero axial charge can evolve  into a dipole distribution along the vorticity direction.  Then the axial charge can exist locally.  If we assume the axial charge density or axial chemical potential $\bar\mu_5$ is small,  we can neglect the axial chemical potential in the second term in Eq.(\ref{Jmu-1-a}) because this term is the first order
contribution, but we should keep the linear term of $\bar\mu_5$ in the first term  because it is actually from  the zeroth order contribution.

 Now let us consider how the axial charge can be separated and redistributed by using the relativistic hydrodynamics.
Since the axial charge is zero initially, we can deal with the evolution of the axial charge in the background of the relativistic uncharged hydrodynamics.
When there is no conserving charge, relativistic hydrodynamic equation is just energy-momentum conservation,
\begin{eqnarray}
\label{div-Tmn}
\partial_\nu T^{\mu\nu}&=&0,
\end{eqnarray}
where $T^{\mu\nu}$ is the energy-momentum tensor and the constitutive equation is given by
\begin{eqnarray}
\label{Tmunu}
 T^{\mu\nu}&=&\left(\varepsilon+P\right) u^\mu u^\nu - P g^{\mu\nu}  +  \pi^{\mu\nu},
\end{eqnarray}
where $\varepsilon$ is the energy density, $  P$ is the pressure of the
fluid and the symmetric tensor $\pi^{\mu\nu}$ is all possible dissipative terms such as shear tensor or bulk tensor and so on.
In Landau frame, the dissipative tensor $\pi^{\mu\nu}$ is orthogonal to the fluid velocity, i.e.,  $\pi^{\mu\nu}u_\nu =\pi^{\mu\nu}u_\mu = 0$. From the  hydrodynamic equation (\ref{div-Tmn}),  we can obtain the following equation directly,
\begin{eqnarray}
\label{div-Tmn-e-1}
u^\nu \partial_\nu \left(T u^\mu\right) -\partial^\mu { T} &=& - \frac{1}{s} \Delta^{\mu\alpha}\partial^\nu \pi_{\alpha\nu},
\end{eqnarray}
where $s$ is entropy density in fluid comoving frame. With the definition of temperature vorticity in Eq.(\ref{div-Tmn}), this equation  implies
\begin{eqnarray}
\label{T-EM-2}
\Omega^{\nu\mu}u_\nu= - \frac{1}{s}\Delta^{\mu\alpha} \partial^\nu \pi_{\alpha\nu} .
\end{eqnarray}
When there is no  electromagnetic field imposed on the chiral system, the axial current is conserved
\begin{eqnarray}
\label{div-j5}
\partial_\mu j_5^\mu =0.
\end{eqnarray}
From the well-known constitutive equation of axial current \cite{Landsteiner:2011cp,Gao:2012ix}
\begin{eqnarray}
\label{j5-mu}
j_5^\mu = n_5 u^\mu + \frac{1}{2}\xi_5 \epsilon^{\mu\nu\rho\sigma}u_\nu \partial_\rho u_\sigma,
\end{eqnarray}
where for the free fermion system the axial charge density $n_5$ and anomalous transport coefficient $\xi_5$ are given by
\begin{eqnarray}
n_5 =\frac{1}{3}\mu_5 T^2,\ \ \ \  \xi_5 = \frac{1}{6}T^2,
\end{eqnarray}
where we have neglected the high order term of $\mu_5$. Substituting Eq.(\ref{j5-mu}) into Eq. (\ref{div-j5}) gives rise to
\begin{eqnarray}
\label{anomaly}
 \partial_\mu (n_5 u^\mu)
=-\frac{1}{24}\Omega_{\mu\nu}\tilde\Omega^{\mu\nu}.
\end{eqnarray}
It is very interesting to  note that the equation (\ref{anomaly}) is very similar to the chiral anomaly of the axial current induced from the electromagnetic field in QED, with the electromagnetic field tensor replaced by the temperature vorticity tensor up to a constant factor.
By using Eq.(\ref{T-EM-2}), we obtain
\begin{eqnarray}
\label{anomaly-1}
\partial_\mu (n_5 u^\mu)
&=&- \frac{1}{6 s}\omega_{T\mu} \partial_\nu \pi^{\mu\nu}\nonumber\\
&\approx&-\frac{1}{3 s}\omega_{T\mu}\Delta_{\nu\lambda}\left(\partial^\lambda \pi^{\mu\nu}
- \pi^{\mu\nu}\partial^\lambda \ln T\right),\ \ \ \ \
\end{eqnarray}
where we have introduced the temperature vorticity vector $\omega_{T\mu}=\tilde\Omega_{\mu\nu}u^\nu$.
In order to arrive at the final equation in Eq.(\ref{anomaly-1}), we have used Eq.(\ref{div-Tmn-e-1}) again and neglected
the dissipative terms.
This result shows that the redistribution of axial charge in uncharged fluid only  happens  for viscous fluid  from initial zero axial charge.
The redistribution can be generated from  the coupling from the spatial vorticity  and spatial gradient of dissipative tensor
or from the spatial vorticity, spatial gradient of temperature and dissipative tensor.

When the axial charge is redistributed,  the first term in Eq.(\ref{Jmu-1-a}) will lead to  extra contribution for  local helicity polarization as
\begin{eqnarray}
\label{S-mu-2}
\Delta S^h(p_T,Y,\phi) &\approx& \frac{\int d\Sigma_\alpha p^\alpha  \bar\mu_5  f (1-f)}
{\int d\Sigma_\alpha p^\alpha f }.
\end{eqnarray}
Hence the final polarization depends on the final distribution of the axial chemical potential and this  can be determined by
the relativistic hydrodynamical simulation.
From the axial charge redistribution, we can also obtain the  extra contribution to the spin polarization along $x$, $y$ and $z$ axis, respectively,
\begin{eqnarray}
\label{S-mu-2}
\Delta S^x (p_T,Y,\phi) &=& \frac{\cos\phi}{\cosh Y}\Delta S^h(p_T,Y,\phi) ,\nonumber\\
\Delta S^y (p_T,Y,\phi) &=& \frac{\sin\phi}{\cosh Y} \Delta S^h(p_T,Y,\phi) ,\nonumber\\
\Delta S^z (p_T,Y,\phi) &=& \tanh Y \Delta S^h(p_T,Y,\phi),
\end{eqnarray}
where $\phi$ is the azimuthal angle of the particle.  It should be noted that
in Refs. \cite{Sun:2018bjl,Liu:2019krs}, the authors simulated the  spin polarization with axial charge redistribution
from the chiral kinetic equation and found that the quark local spin polarizations exhibit  an azimuthal asymmetry similar to
the experimental data for the $\Lambda$ hyperon.  Then it will be very valuable to investigate the spin polarization from the helicity charge
redistribution in the scenario of  the relativistic hydrodynamics.

\section{Helicity  correlation }

In previous discussion, we only restricted ourselves to the axial charge induced locally by redistribution from initial zero value.  In the heavy ion collisions at very high energy,
the net helicity could exist at the beginning due to  classical color longitudinal fields  just after the collision \cite{Kharzeev:2001ev,Lappi:2006fp}
or QCD sphaleron  transitions   in the quark gluon plasma \cite{McLerran:1990de,Moore:1997im,Moore:1999fs,Bodeker:1999gx,Shuryak:2002qz}.
Such net initial axial charge could lead to the well-known chiral magnetic effect \cite{Vilenkin:1980fu,Kharzeev:2007jp,Fukushima:2008xe} and result in electric charge separation along the angular momentum in non-central heavy ion collisions. Because the net initial axial charge with positive and negative sign should be produced with equal probability in
many events, the helicity polarization of one  particle after averaging  over these different  events will vanish. However we can detect this net axial charge  by measuring the
helicity correlation of two hyperons event-by-evently \cite{Pang:2016igs,Becattini:2020xbh}.

As we all know, the polarization of hyperons can be determined from the angular distribution of hyperon decay products in the hyperon rest frame
\begin{eqnarray}
\label{Hyperon-decay}
\frac{dN}{d\cos\theta^*} &=& \frac{1}{2}\left(1+ \alpha_H P_H^h \cos\theta^*\right).
\end{eqnarray}
where  $\theta^*$  is the angle between the polarization direction (here it is just  hyperon's momentum in lab frame) and the momentum of hyperon's decay products 
in hyperon's rest frame and $P_H^h$ denotes the helicity polarization of hyperon $H$ .  Now we choose two hyperons $H_1$ and $H_2$ which must be in  one same event and calculate the average value event-by-evently,
\begin{eqnarray}
\label{correlation}
C_{H_1 H_2}&\equiv&\left\langle\frac{dN_1}{d\cos\theta_1^*}\frac{dN_1}{d\cos\theta_2^*}\cos\theta_1^* \cos\theta_2^*\right\rangle \nonumber\\
&=& \frac{1}{4}\alpha_{H_1}\alpha_{H_2}  P_{H_1}^h P_{H_2}^h.
\end{eqnarray}
If the two hyperons are the same, then the correlation will be given by
\begin{eqnarray}
C_{HH}=\frac{1}{4}\alpha_{H}^2 \left(P_{H}^{h}\right)^2,
\end{eqnarray}
which is positive. If the two hyperons are  particle and antiparticle and assume the polarization for them are the same, then the correlation will be given by
\begin{eqnarray}
C_{H\bar H}=-\frac{1}{4}\alpha_{H}^2 \left(P_{H}^{h}\right)^2,
\end{eqnarray}
which is negative. In measuring such initial helicity polarization, we do not need to determine the reaction plane and we can even detect it in central heavy ion collisions.
After taking an average over many events without determining the reaction plane, we expect all the spin polarization from the helicity charge redistribution will be cancelled and only initial net helicity charge survives. Helicity polarization correlation measures the global helicity polarization.

\section{Helicity  alignment}
Similarly, we can measure the spin alignment along the direction of the vector meson's momentum, which can be denoted by helicity
alignment. Spin alignment for vector mesons can be described by a Hermitian $3\times 3$ spin-density matrix $\rho$. The matrix element
$\rho_{00}$ can be determined from the angular distribution of the decay products\cite{Schilling:1969um},
\begin{eqnarray}
\label{Meson-decay}
\frac{dN}{d\cos\theta^*} &=&\frac{3}{4}
\left[ 1 - \rho_{00}+(3\rho_{00}-1)\cos^2\theta^*\right],
\end{eqnarray}
where we take the vector meson's momentum direction $\hat{\bf p}=(\sin\theta \cos\phi, \sin\theta \sin\phi, \cos\theta)$ as the quantization axis.
When  $\rho_{00}$ deviates from $1/3$,   spin alignment will arise.
We assume that the spin density matrix for quarks or antiquarks is diagonal as \cite{Liang:2004xn}
\begin{eqnarray}
\label{density-qurk}
\rho_{q/\bar q}^h&=& \frac{1}{2}
\left(
  \begin{array}{cc}
    1+ P^h_{q/\bar q} & 0 \\
    0 & 1-P^h_{q/\bar q} \\
  \end{array}
\right),
\end{eqnarray}
where $P^h_{q/\bar q}$ denotes the polarization for quarks or antiquarks along the particle's momentum.
In the recombination scenario, we assume that the quark and antiquark combine into a vector meson only when their momentum is
along the similar direction. For simpicity, we will set the quark and antiquark in the same direction.
Then we can obtain the density matrix for vector meson from Eq.(\ref{density-qurk}) \cite{Liang:2004xn}
\begin{eqnarray}
\label{density-vector}
\rho&=&\left(
      \begin{array}{cccc}
      \rho_{11} & 0 & 0  \\
        0 & \rho_{00}  & 0 \\
        0 & 0  &   \rho_{-1-1} \\
      \end{array}
    \right),
\end{eqnarray}
where
\begin{eqnarray}
\rho_{11} &=& \frac{ (1+P_q^h) (1+P_{\bar q}^h) }{3+ P_q^h P_{\bar q}^h},\\
\rho_{00} &=& \frac{ 1 - P_q^h P_{\bar q}^h }{3+ P_q^h P_{\bar q}^h},\\
\rho_{-1-1} &=& \frac{ (1-P_q^h) (1-P_{\bar q}^h) }{3+ P_q^h P_{\bar q}^h}.
\end{eqnarray}
Then the deviation from $1/3$ for $\rho_{00}$ is given by
\begin{eqnarray}
\Delta\equiv\rho_{00}-\frac{1}{3}=\frac{1-P_q^h P_{\bar q}^h}{3+ P_q^h P_{\bar q}^h}-\frac{1}{3}
\approx -\frac{4}{9}P_q^h P_{\bar q}^h,
\end{eqnarray}
when the polarization of quark or anti-quark is very small.
Given the helicity density matrix, we can calculate the spin density matrix along any spin-quantization axis and obtain the spin alignment
along these different directions.
We use $\rho^x$, $\rho^y$ and $\rho^z$ to denote the density matrices when we quantize the spin in $x$, $y$ and $z$ axis, respectively.
The matrix elements  $\rho^x_{00}$,$\rho^y_{00}$  and $\rho^z_{00}$  are given by
\begin{eqnarray}
\label{rho-x}
\rho_{00}^x &=&\frac{1-\rho_{00}}{2} + \frac{3\rho_{00}-1}{2}\sin^2\theta \cos^2\phi,\\
\label{rho-y}
\rho_{00}^y&=&\frac{1-\rho_{00}}{2} + \frac{3\rho_{00}-1}{2}\sin^2\theta \sin^2\phi,\\
\label{rho-z}
\rho_{00}^z &=&  \rho_{00}  + \frac{1-3\rho_{00}}{2}\sin^2\theta.
\end{eqnarray}
We note that
\begin{eqnarray}
\rho_{00}^x + \rho_{00}^y + \rho_{00}^z &=&  1.
\end{eqnarray}
Actually this identity holds for any normalized spin-1 density matrix, i.e., the  sum of  the 00-components of normalized spin density matrices
 in three orthogonal spin-quantization axis must be unit.
 With the diagonal density matrix (\ref{density-vector}) and the results (\ref{rho-x}-\ref{rho-z}), we can easily  obtain the $00$-component of the transverse  density matrix along any direction orthogonal to the particle's momentum
\begin{eqnarray}
\rho_{00}^T &=&  \frac{1-\rho_{00}}{2}.
\end{eqnarray}
In heavy ion collision, we can choose this transverse direction as the  normal vector of the reaction plane.

If we assume  the momentum distribution of the vector meson in the form
\begin{eqnarray}
E\frac{dN}{d^3 p}= N_0 \left(1 + 2 v_2 \cos 2\phi\right)
\end{eqnarray}
where only the elliptic flow $v_2$ is  retained. Then the averaged  $\rho^{x}_{00}$, $\rho^{y}_{00}$ and $\rho^{z}_{00}$ at $Y=0$ are  given by, respectively,
\begin{eqnarray}
\bar\rho^{x}_{00} &=& \frac{1}{3} +\frac{1}{4}\Delta +\frac{3}{4}v_2 \Delta ,\\
\bar\rho^{y}_{00} &=& \frac{1}{3} +\frac{1}{4}\Delta -\frac{3}{4}v_2 \Delta ,\\
\bar\rho_{00}^z  &=&  \frac{1}{3}-\frac{1}{2}\Delta
\end{eqnarray}
These results explicitly show how the helicity alignment from helicity charge can result in the  spin alignment along the transverse or longitudinal directions.

Similar to the polarization correlation, the deviation is also proportional to
 the square of the quark's polarization. However the difference between them is
 that the helicity polarization  correlation must be from the average or net
 global helicity charge, but the spin alignment could be from the local polarization or  event-by-event fluctuation  \cite{Sheng:2020ghv,Xia:2020tyd}.
The relevant quantity  for hyperon polarization correlation is the square of mean polarization $\langle P_{q/\bar q} \rangle^2$ while
the relevant quantity for spin alignment is the average of the square of polarization $\langle P_{q/\bar q}^2 \rangle$.
Hence in general, the spin alignment should be larger than the  polarization  for the hyperon. Because of this local polarization effect or fluctuation, it is hard to detect the initial net helicity charge by helicity alignment even in central heavy ion collisions if there exists
strong topological charge fluctuation.  The strong alignment measured by STAR
and ALICE which is not consistent with hyperon polarization might be due to the fact that the global polarization is very small while
 local polarization fluctuation is very large.

\section{Summary and outlook}

We have discussed the possible spin polarization along the direction of the detected hadrons---helicity polarization. We find that the vorticity produced in non-cental heavy-ion
collisions can also lead to the helicity polarization, which could emerge locally with azimuthal asymmetry and rapidity odd dependence. The viscous hydrodynamics could induce
the temperature vorticity from initial zero value and lead to the helicity charge redistribution. Such helicity  charge redistribution can  result in helicity polarization and longitudinal or transverse  polarization as well. When the topological charge or initial color longitudinal field  fluctuates event-by-evently, the global net helicity charge could be measured by the correlation of two hyperon's helicity
polarization  and local helicity charge fluctuation could be measured by  the spin alignment along the vector meson's momentum in central heavy ion collisions at very high energies.

\section{Acknowledgments}

The author thanks Ai-Hong Tang for fruitful discussion. This work was supported in part by the National Natural Science Foundation of China under
Nos. 11890710, 11890713 and 11475104 and the Natural Science Foundation of Shandong Province under No. JQ201601.

\end{document}